%% file: HIN-21-009_temp.tex
\pdfoutput=1
\documentclass[11pt,twoside,a4paper,cmspaper,final,collab]{cms-tdr}
\def\svnVersion{702cb1b}\def\svnDate{2023/10/18}\def\cmsCernNoTag{CERN-EP-2022-098}\def\cmsCernDate{\today}\def\cmsMessage{Published in Physical Review Letters as \href{http://dx.doi.org/10.1103/PhysRevLett.131.151803}{\doi{10.1103/PhysRevLett.131.151803}.}}
\begin{document}\cmsNoteHeader{HIN-21-009}

\newcommand{\PbPb}{\ensuremath{\mathrm{Pb}\mathrm{Pb}}\xspace}
\newcommand{\tautau}{\ensuremath{\PGtp{}\PGtm{}}\xspace}
\newcommand{\nch}{\ensuremath{n_\text{ch}}\xspace}
\newcommand{\result}{\ensuremath{4.8\pm 0.6\pm 0.5,\mu\text{b}}\xspace}
\newcommand{\xsecresult}{\ensuremath{\xsec=\result}\xspace}
\newcommand{\limit}{\ensuremath{0.001^{+0.055}_{-0.089}}\xspace}
\newcommand{\atau}{\ensuremath{a_{\PGt}}\xspace}
\newcommand{\threeprong}{\ensuremath{\PGt_{\text{3prong}}}\xspace}
\newcommand{\muonic}{\ensuremath{\PGt_{\PGm}}\xspace}
\newcommand{\gammagammatautau}{\ensuremath{\PGg{}\PGg\to\tautau}\xspace}
\newcommand{\gammagammaccbar}{\ensuremath{\PGg{}\PGg\to\PQc{}\PAQc}\xspace}
\newcommand{\gammagammabbbar}{\ensuremath{\PGg{}\PGg\to\PQb{}\PAQb}\xspace}
\newcommand{\xsec}{\ensuremath{\sigma(\gammagammatautau)}\xspace}
\newcommand{\nsigtautau}{\ensuremath{N_{\text{sig}}}\xspace}
\newcommand{\ptvis}{\ensuremath{\pt^{\text{vis}}}\xspace}
\newcommand{\mvis}{\ensuremath{m_{\PGt}^{\text{vis}}}\xspace}
\newcommand{\Dphi}{\ensuremath{\Delta\phi(\muonic,\threeprong)}\xspace}
\newcommand{\Bmuonic}{\ensuremath{\mathcal{B}_{\muonic}}\xspace}
\newcommand{\Bthreeprong}{\ensuremath{\mathcal{B}_{\threeprong}}\xspace}

\newlength\cmsTabSkip\setlength{\cmsTabSkip}{1ex}

\cmsNoteHeader{HIN-21-009}
\title{Observation of \texorpdfstring{\PGt}{tau} lepton pair production in ultraperipheral lead-lead collisions at \texorpdfstring{$\sqrtsNN = 5.02\TeV$}{sqrt(sNN) = 5.02 TeV}}

\date{\today}

\abstract{We present an observation of photon-photon production of \PGt lepton pairs in ultraperipheral lead-lead collisions. The measurement is based on a data sample with an integrated luminosity of 404\mubinv collected by the CMS experiment at a center-of-mass energy per nucleon pair of $\sqrtsNN = 5.02\TeV$. The \gammagammatautau process is observed for \tautau events with a muon and three charged hadrons in the final state. The measured fiducial cross section is $\xsec = \result$, where the second (third) term corresponds to the statistical (systematic) uncertainty in \xsec, in agreement with leading-order QED predictions. Using \xsec, we estimate a model-dependent value of the anomalous magnetic moment of the \PGt lepton of $\atau=\limit$. 
}

\hypersetup{%
pdfauthor={CMS Collaboration},%
pdftitle={Observation of tau lepton pair production in ultraperipheral lead-lead collisions at sqrt(s[NN])=5.02 TeV},%
pdfsubject={CMS},
pdfkeywords={CMS, heavy ions, tau lepton}}

\maketitle

Ultraperipheral collisions (UPCs) of nuclei, where the impact parameter is larger than the sum of the nuclear radii, provide an extremely clean environment to study various photon-induced processes~\cite{Baltz:2007kq}. For the case of lead-lead (\PbPb) UPCs, the production cross section for two-photon fusion processes is enhanced by a factor of about $Z^4$ (where $Z=82$ is the Pb charge number), relative to proton-proton collisions. The possibility of observing photon-induced \PGt lepton production in UPC events at a heavy ion collider was considered well before the LHC era~\cite{delAguila:1991rm}. Recently, theoretical studies~\cite{Beresford:2019gww,Dyndal:2020yen} have proposed that kinematic properties of \PGt lepton pairs produced in heavy ion UPCs at the LHC can be used to constrain the electromagnetic couplings of the \PGt lepton. These constraints allow for fundamental tests of quantum electrodynamics (QED) and for probing beyond the standard model (BSM) physics. 

A contributing factor in the coupling of the lepton (\Pell) to the photon (\PGg) is the anomalous magnetic moment $a_{\Pell} = (g-2)_{\Pell} / 2$, with the $g$-factor being the proportionality constant that relates the magnetic moment to the spin of the lepton. With 12 significant digits, the electron anomalous magnetic moment  $a_{\Pe}$ is among the most precisely measured quantities~\cite{Parker:2018vye}, and differs from the standard model (SM) expectation by either  $-2.4$ or $+1.6$ standard deviations~\cite{Parker:2018vye,Morel:2020dww}, depending on the input value of the fine structure constant, $\alpha_\text{QED}$. The value of $a_{\PGm}$ has been measured to 9 significant figures~\cite{Muong-2:2021ojo}. It shows a tension of $+4.2$ standard deviations with respect to SM predictions~\cite{Aoyama:2020ynm}, although a calculation with a modified hadronic contribution~\cite{Muong-2:lattice} reduces the discrepancy between data and theory by a factor of more than 2, albeit with an uncertainty that is about 20\% larger. While the predicted value of \atau is $0.00117721\,(5)$~\cite{Passera:2006gc,Eidelman:2007sb},  with the number in parentheses denoting the uncertainty in the least significant figure, its best measured value is $-0.018\pm0.017$ from the DELPHI Collaboration~\cite{DELPHI:2003nah} (other existing limits on \atau can be found in Ref.~\cite{ParticleDataGroup:2020ssz}). The larger uncertainty in \atau compared to $a_{\PGm}$ and $a_{\Pe}$ measurements primarily results from the short \PGt lepton lifetime, which is of the order of  $10^{-13}\unit{s}$, such that \PGt leptons cannot be stored long enough to measure their \atau-dependent precession in a magnetic field. A more precise \atau determination would facilitate tighter constraints on BSM physics models~\cite{Crivellin:2021spu,Crivellin:2021rbq}, in which additional particles with mass $M$ contribute with terms typically proportional to $(m_\ell/M)^2$. This motivates employing novel experimental approaches for measuring \atau at current and potential future colliders, as undertaken in this Letter and in a recent measurement by the ATLAS Collaboration~\cite{ATLAS:2022ryk}.

Here, we present an observation of \PGt lepton pairs in ultraperipheral \PbPb collisions, \gammagammatautau,  in events that may contain excitations of the outgoing Pb ions. The analysis is based on a data sample with an integrated luminosity of 404\mubinv collected by the CMS experiment in 2015 at a center-of-mass energy per nucleon pair of $\sqrtsNN = 5.02\TeV$. One \PGt lepton is reconstructed through its decay to one muon and two neutrinos, while the other is reconstructed through its ``3 pronged'' decay into hadrons plus a neutrino~\cite{ParticleDataGroup:2020ssz}. This choice of final state offers a clean experimental signature, with the muon used for online selection and the hadronically decaying \PGt candidate providing discrimination against dimuon photoproduction and an unambiguous reconstruction of \PGt lepton decay. The reconstruction of the \PGt leptons is performed over a fiducial phase space, defined by the transverse momentum (\pt) and pseudorapidity ($\eta$) of each particle. Tabulated results are provided in the HEPData record for this analysis~\cite{HEPData}.

The CMS apparatus~\cite{CMS:2008xjf} is a multipurpose, nearly hermetic detector, designed to trigger on~\cite{CMS:2020cmk,CMS:2016ngn} and identify electrons, photons, muons, \PGt leptons, jets, and missing \pt~\cite{CMS:2021EG,Sirunyan:2018fpa,CMS:2014pgm}. A global reconstruction ``particle-flow'' algorithm~\cite{CMS:2017yfk} combines the information provided by the all-silicon inner tracker, the crystal electromagnetic calorimeter, and the brass and scintillator hadron calorimeter, operating inside a 3.8\unit{T} superconducting solenoid, with that from gas-ionization muon detectors embedded in the flux-return yoke outside the solenoid, to build \PGt lepton candidates and jets, and to measure the missing \pt~\cite{CMS:2018jrd,CMS:2016lmd,CMS:2019ctu}. Forward hadron (HF) calorimeters~\cite{CMS:2019hpr}, made of steel and quartz-fibers, extend the $\abs{\eta}$ coverage from 3.0,  provided by the barrel and endcap detectors, to 5.2.  The HF calorimeters are segmented to form $\Delta\eta{\times}\Delta\phi$ ``towers'' of width $0.175{\times}0.175$, with $\phi$ being the azimuthal angle. Events are selected online using a two-tiered trigger system. The first level, composed of custom hardware processors, uses information from the calorimeters and muon detectors~\cite{CMS:2020cmk}. The second level, known as the high-level trigger~\cite{CMS:2016ngn}, consists of a farm of processors running a version of the full event reconstruction software.

The UPCs producing two final-state \PGt leptons are uniquely characterized by low track multiplicity and the presence of very forward (i.e., high $\abs{\eta}$) lead ions that are either scattered or dissociated in a direction so close to the beam as to be undetectable. Therefore, we select high-purity UPC events~\cite{CMS:2018erd} by requiring in real time the presence of a single muon with no explicit \pt threshold requirement, at least one pixel detector track, and low event activity in the HF~\cite{CMS:2020cmk}. To further suppress background processes, such as hadronic \PbPb collisions, it is required offline that the maximum energy measured in an HF tower is below 4\GeV. 

Furthermore, the fiducial phase space region is constrained offline by selecting events with one muon and exactly three additional tracks. For the muon defining the ``\muonic'' candidate, a selection is applied requiring $\abs{\eta}<2.4$ and that the muon satisfies the ``soft'' identification criteria described in Ref.~\cite{Sirunyan:2018fpa}, with $\pt>3.5\GeV$ for $\abs{\eta}<1.2$ and $\pt>2.5\GeV$ for $\abs{\eta}>1.2$, following the acceptance of the muon detector system. The three tracks that form the ``\threeprong'' candidate~\cite{CMS:2018jrd} are assumed to be pions and are selected within the tracker acceptance ($\abs{\eta}<2.5$). Along the direction of the two beams, they are required to have a common vertex within 2.5\,mm relative to the vertex corresponding to the hardest scattering in the event~\cite{CMS-TDR-15-02}. In addition, they must be identified as charged hadrons by the particle-flow algorithm. The transverse momentum of the leading (\ie, the highest \pt) and two subleading pions must be greater than 0.5 and 0.3\GeV, respectively. The selected tracks are required to pass the ``high-purity'' requirements of Ref.~\cite{CMS:2014pgm}. The \threeprong candidate is then required to be of opposite charge relative to the selected \muonic, and to have $\ptvis>2\GeV$, where \ptvis is the vector sum \pt of the three charged pions (the ``visible'' decay products of the \threeprong candidate). Additionally, the invariant mass of the three pion candidates \mvis is required to be less than 1.5\GeV. 
 With these selections we identify 91 \gammagammatautau candidate events.
 
\sloppy Backgrounds arise from heavy quark photoproduction, UPC photon-photon and photon-pomeron interactions producing mesons that can decay to muons and charged hadrons. Dedicated samples of events from \gammagammatautau~\cite{Beresford:2019gww}, \gammagammaccbar, and \gammagammabbbar processes are generated with \MGvATNLO(v2.6.5)~\cite{Alwall:2014hca}, where \PYTHIA{}8~(v2.1.2)~\cite{Sjostrand:2014zea} is used for the hadronization and decay, and \GEANTfour~\cite{Agostinelli:2002hh} is used to emulate
the full CMS detector response.
 All studied kinematic distributions of the muons and charged pions in simulated events are corrected using comparisons between the simulation and data, outside the signal region, as a function of the muon or track \pt and $\eta$. For muons, we use a ``tag-and-probe'' method with $\JPsi\to\PGmp\PGmm$ events~\cite{Sirunyan:2018fpa}. For charged hadrons, we use the number of reconstructed \PDz meson decays to final states with four charged hadrons divided by those with two daughters.
The simulated background processes produce a large number of tracks and hence sparsely populate the signal-dominated phase space region. They are only used to partly validate the expected \gammagammaccbar and \gammagammabbbar contributions to the background estimation as described in the following paragraph.

\begin{figure*}[htbp!]
\centering
\includegraphics[width=0.32\textwidth]{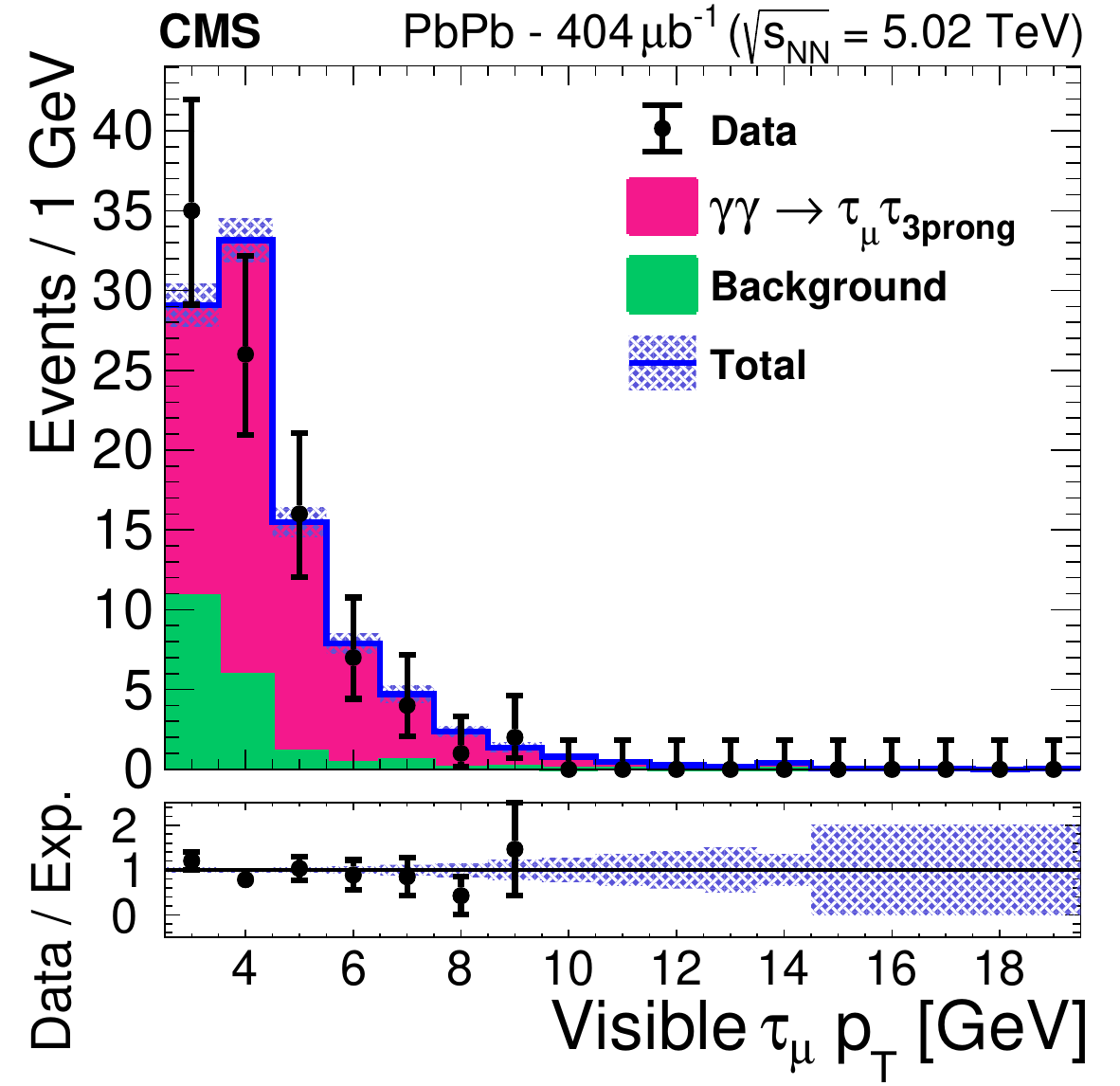}
\includegraphics[width=0.32\textwidth]{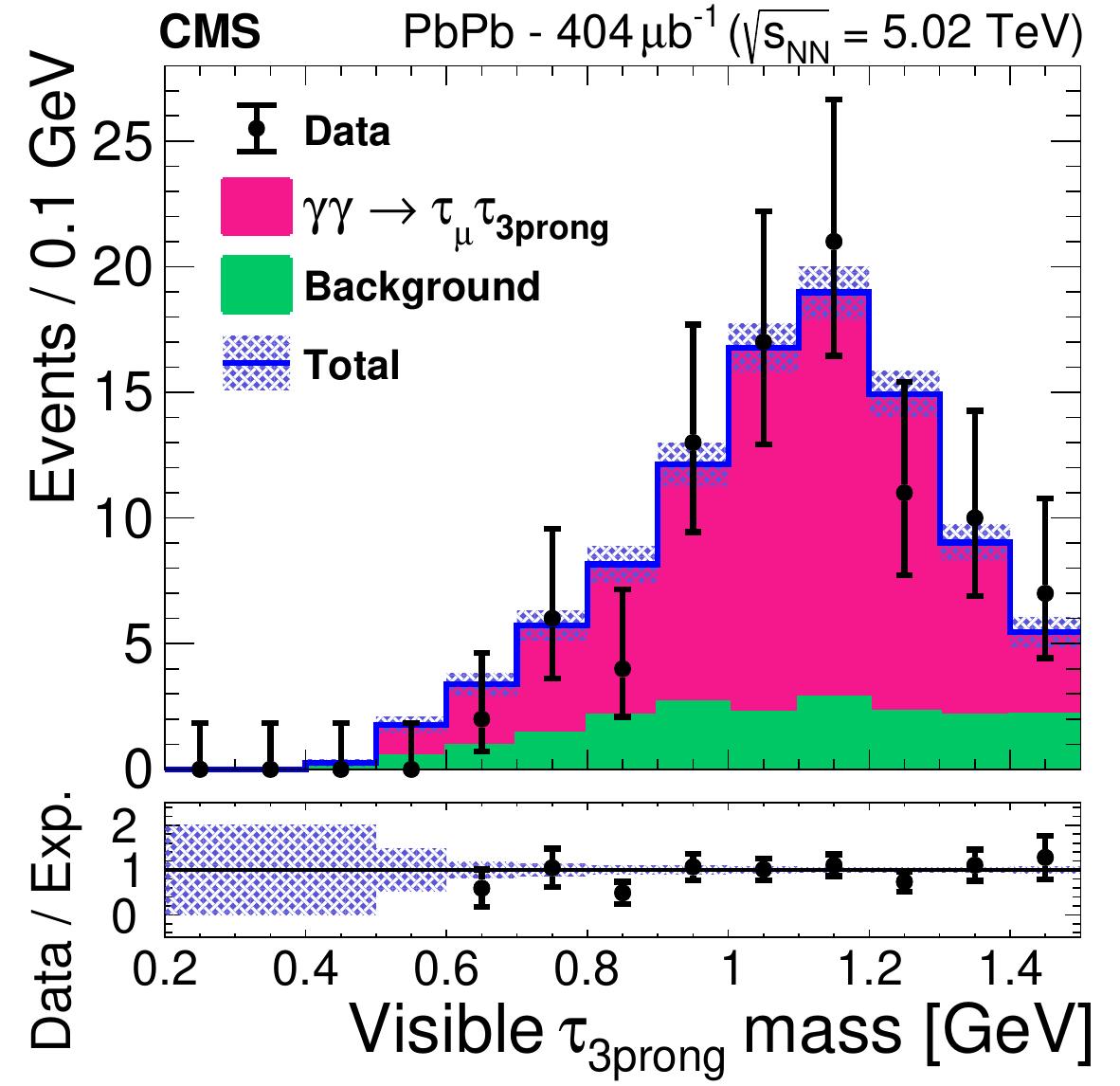}
\includegraphics[width=0.32\textwidth]{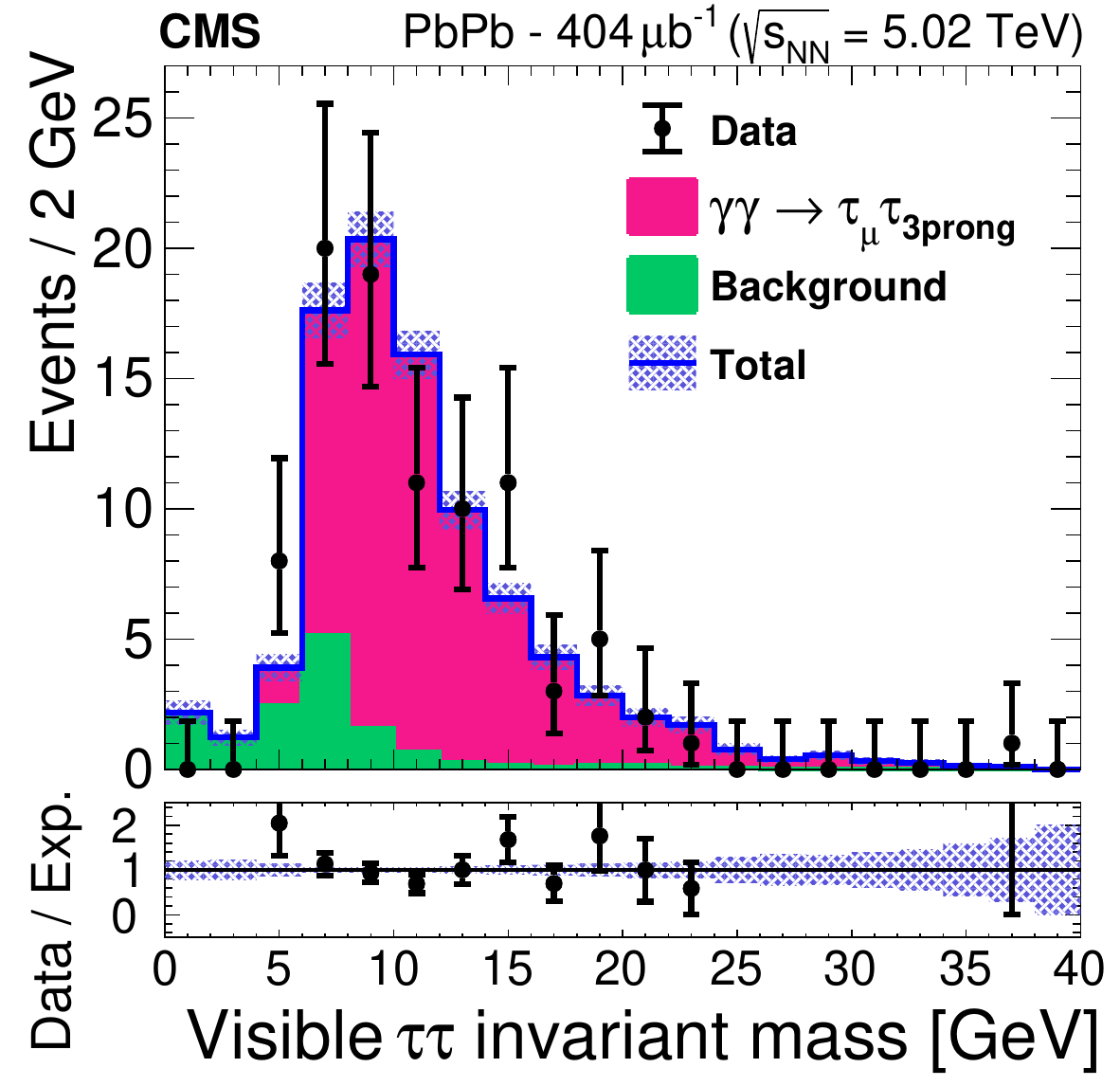}
\caption{Left: Transverse momentum of the muon originating from the \muonic candidate. Middle: Invariant mass of the three pions forming the \threeprong candidate. Right: \tautau invariant mass. In all plots, the signal component (magenta histogram) is stacked on top of the background component (green histogram), considering their initial normalizations, as described in the text. The sum of signal and background is displayed by a blue line and the shaded area shows the statistical uncertainty. The data are represented with black points and the uncertainty is statistical only. The lower panels show the ratios of data to the signal-plus-background prediction, and the shaded bands represent the statistical uncertainty in the prefit expectation.}
\label{fig:muonKinematicsDataMC}
\end{figure*}

To properly estimate the background, we use a technique based on control samples in data, referred to as the ``ABCD method''. Three phase space regions (``categories'') are used to derive the background in the fourth region,  from which the signal is extracted. The four categories, which have been found to be uncorrelated in data, are defined according to the value of the highest energy tower in HF, and the number of charged particle tracks per event (\nch), excluding the track associated with the \muonic candidate. The low-\nch categories (B and D) are defined by $\nch=3$, whereas the high-\nch categories (A and C) must have $5\leq \nch\leq8$ to avoid signal contamination while being similar to the signal region. The low-HF (C and D) and high-HF (A and B) categories are defined by energies below and above 4\GeV, respectively. Consequently, category D is the signal region (low-\nch and low-HF category), and the background estimation is $\text{B}_i\,\text{C}_i/\text{A}_i$, where each of the categories is evaluated per kinematic-variable and category-dependent bin, as indicated by the subscript $i$. Based on the simulated signal events, we find that the event selection described above removes all signal events from the control regions (A--C).
The kinematic distributions showing the \gammagammatautau signal process, scaled to match the QED prediction of Ref.~\cite{Beresford:2019gww}, as well as the background model based on control samples in data, are shown in Fig.~\ref{fig:muonKinematicsDataMC}. Good agreement is observed between the measured distributions and the sum of the signal simulation and background estimation.

A binned maximum likelihood fit of signal and background components is used for the signal extraction. The fit is performed on the binned distribution of the difference in azimuthal opening angle between the \muonic and \threeprong candidates, \Dphi, exploiting the fact that the two signal \PGt leptons are produced azimuthally back-to-back in UPCs~\cite{Baltz:2007kq,CMS:2020skx}. The signal distribution is derived from the \gammagammatautau simulation, while that of the background is obtained from the ABCD method described above, including its normalization as a constant parameter in the fit. The initial (``prefit'') number of signal events is taken from the QED prediction of Ref.~\cite{Beresford:2019gww}. Systematic uncertainties may affect both the normalization and the shape of the \Dphi distributions. These uncertainties, in addition to the bin-by-bin variations of the signal and background templates, are represented by nuisance parameters in the fit. Rate-changing nuisance parameters are represented as log-normal probability distribution functions, while shape-changing ones are represented with Gaussian probability distribution functions. The negative of the log likelihood is minimized by varying the nuisance parameters according to their uncertainties and by scaling the signal by a multiplicative factor $r$.

Uncertainties arising from the HF energy threshold are evaluated by varying the HF energy by 10\%~\cite{CMS:2018bbk}. The effect on the measured cross section due to this variation is dominated by the resulting variation in the background shape from the ABCD procedure, and is found to be 0.9\%. An additional systematic uncertainty coming from the background shape and yield estimation is considered by reevaluating the background using the ABCD procedure, changing the high \nch parameter to individual values of 5, 6, 7, and 8, as opposed to the range 5--8. The maximum variation with respect to the central value comes from the determination with $\nch=5$, resulting in a 0.2\% variation of the fiducial cross section measurement.

The uncertainty in the muon efficiency, including the trigger response, identification and tracking efficiency, has an impact of 6.7\%. The integrated luminosity is measured with the methods described in Refs.~\cite{CMS:2021xjt,CMS-PAS-LUM-18-001}, and has an uncertainty of 5\%, which affects the yield from the QED simulation to which the signal is normalized. The uncertainty in the pion tracking efficiency results in an uncertainty of 3.6\%. The simulated signal distribution has a finite number of events, resulting in a 3\% uncertainty due to bin-by-bin statistical fluctuations, and a 1.1\% weighted binomial uncertainty on the efficiency. The uncertainty in the \PGt lepton branching fraction measurements is 0.6\%~\cite{ParticleDataGroup:2020ssz}.

The total uncertainty, obtained by adding them in quadrature while taking into account their correlation, is found to be 9.7\%.

\begin{figure}[htbp!]
\centering
\includegraphics[width=0.5\textwidth]{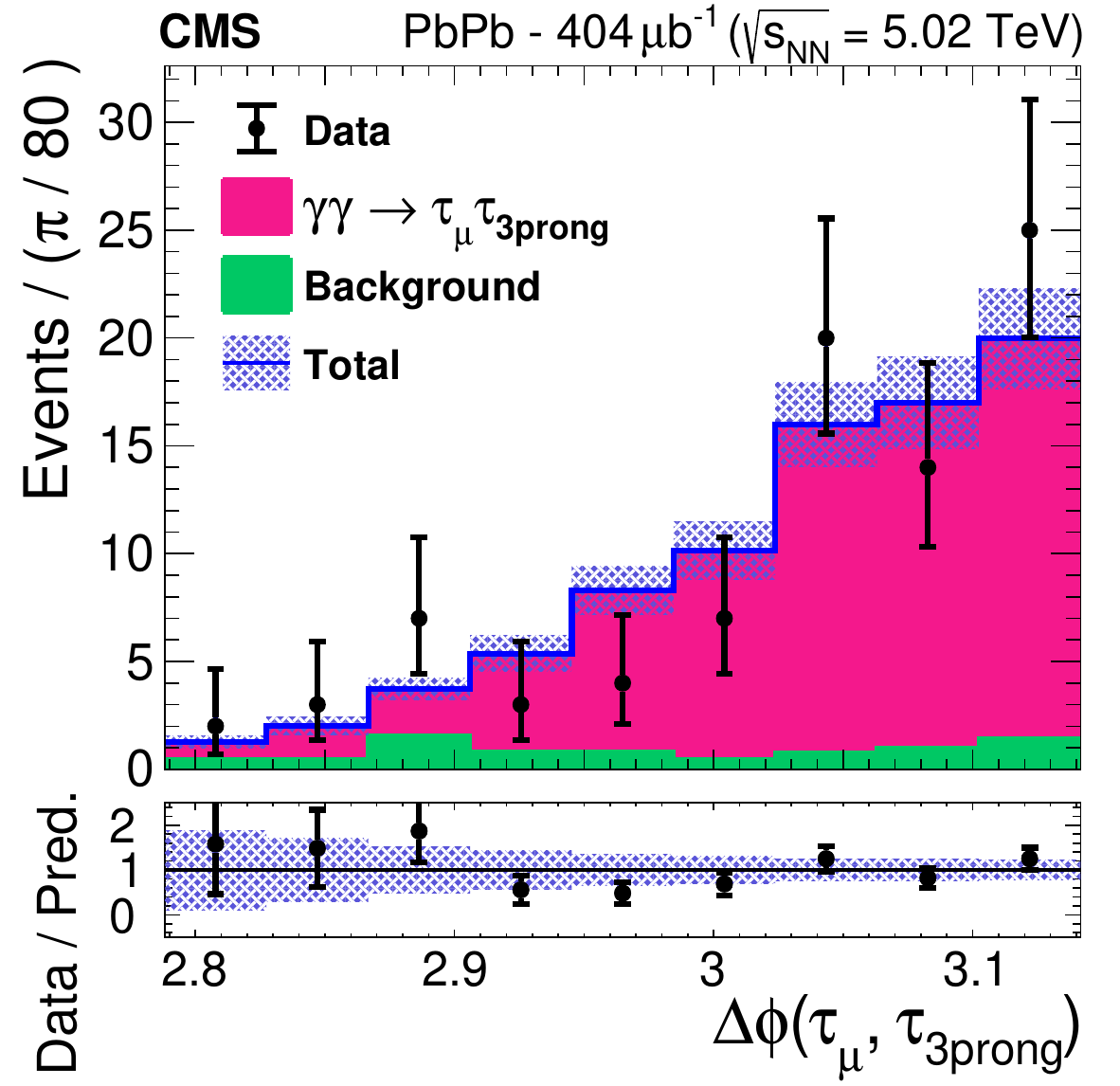}
\caption{Difference in azimuthal opening angle between the \muonic and $\threeprong$ candidates. The data are represented by the points with the vertical bars showing the statistical uncertainties. The signal (background) contribution is given by the magenta (green) histogram, after the application of the fit procedure. The total is displayed by a blue line and the shaded area shows the combined statistical and systematic uncertainties. The lower panel shows the ratio of data to the signal plus background prediction, and the shaded band represents the total uncertainty in the postfit prediction.}
\label{fig:postfit}
\end{figure}

The best fit value of the signal strength multiplicative factor is $r=0.99^{+0.16}_{-0.14}$ with $\nsigtautau=77\pm 12$ signal events in the integral of the postfit signal component. The fit result is shown in Fig.~\ref{fig:postfit}, along with the data, and signal and background templates. The observed (expected) signal significance, computed using the asymptotic approximation~\cite{Cowan:2010js}, is found to be 14.2 (14.5) standard deviations. These values indicate a clear observation of the \gammagammatautau process.

The cross section is measured in the fiducial phase space region, following the kinematic requirements previously described.
 The formula used is \xsec = $\nsigtautau/(2\epsilon\,\Lumi_\text{int}\,\Bmuonic\,\Bthreeprong)$, where \nsigtautau is the number of signal events estimated by the fit process, $\epsilon$ is the total signal efficiency, $\Lumi_\text{int}=404\pm 20\mubinv$ is the total integrated luminosity, and $\Bmuonic=(17.39\pm 0.04)$\% and $\Bthreeprong=(14.55\pm 0.06)$\%~\cite{ParticleDataGroup:2020ssz} are the branching fractions for the two \PGt lepton decay modes. The factor of two accounts for the two potential \PGt lepton decay combinations yielding the same final state, whereas 3-prong decays could include additional neutral pions.  The efficiency is the product of the pion and muon reconstruction, the trigger, and the analysis selection efficiencies, and is evaluated using simulated signal events. The efficiency is calculated as the number of reconstructed events passing the analysis selection criteria divided by the number of generated events inside the fiducial phase space region, and is found to be $\epsilon= (78.5\pm 0.8)$\%.

\begin{figure}[htbp!]
  \centering
    \includegraphics[width=0.48\textwidth]{Figure_003.pdf}
    \caption{The cross section, \xsec, along with its statistical \stat and systematic \syst uncertainties, as measured in a fiducial phase space region at $\sqrtsNN = 5.02\TeV$. The theoretical predictions~\cite{Beresford:2019gww,Dyndal:2020yen} are computed with leading-order accuracy in QED and are represented by the vertical solid lines that can be compared with the vertical dotted line representing this measurement. The outer blue (inner red) error bars surrounding the data point indicate the total (statistical) uncertainties, whereas the green hatched bands correspond to the uncertainty in the theoretical predictions as described in the main text. The potential electromagnetic excitation of the outgoing Pb ions is denoted by $(^{\ast})$.}
    \label{fig:sum}
\end{figure}

Combining all of the above, the fiducial cross section is found to be \xsecresult, where the second (third) term corresponds to the statistical (systematic) uncertainty in \xsec. The result, summarized in Fig.~\ref{fig:sum}, is compared to leading-order QED predictions~\cite{Beresford:2019gww,Dyndal:2020yen}. The analytical calculation from Ref.~\cite{Dyndal:2020yen} results in a cross section which is 20\% higher than that found in Ref.~\cite{Beresford:2019gww}. This is explained in Ref.~\cite{Dyndal:2020yen} as mainly stemming from the different requirements applied in the modeling of single-photon fluxes. In both cases, although further theory advancements are needed for a proper uncertainty evaluation, a conservative uncertainty of 10\% is considered following the approach from Ref.~\cite{CMS:2018erd} given the similarity of final states and phase-space volumes.

Recent calculations have evaluated the impact of BSM processes on the \gammagammatautau cross section. The BSM coupling variations in \atau can change the expected cross section and alter the \PGt lepton \pt spectrum~\cite{Beresford:2019gww,Dyndal:2020yen}. We assume the correction factor of Ref.~\cite{Beresford:2019gww} to extrapolate the fiducial cross section measurement
 to the full phase space region, after taking into account an extra factor of $1/\sqrt{4\pi}$ for the electron charge in Heaviside-Lorentz units. We then use the dependency of the total \xsec as a function of \atau~\cite{Beresford:2019gww} to extract a model-dependent value of \atau at the LHC. The measured value is $\atau=\limit$, which is consistent with the current best measurement~\cite{DELPHI:2003nah}. The ATLAS Collaboration has also recently reported a measurement of \gammagammatautau using a larger \PbPb data sample with an integrated luminosity of 1.44\nbinv~\cite{ATLAS:2022ryk}.  With respect to the ATLAS measurement, we cover a larger phase space with muon $\pt>2.5\GeV$, while Ref.~\cite{ATLAS:2022ryk} uses $\pt>4\GeV$, and we make no restrictions on neutron emission. Because of the larger fiducial phase space region comprised by our measurement, the attained precision in $r$ for the studied final state is comparable to that of $r=0.98^{+0.14}_{-0.13}$ obtained in Ref.~\cite{ATLAS:2022ryk}. The approaches followed by the two collaborations in the measurement of \atau are complementary to each other: we extract \atau from \xsec, while Ref.~\cite{ATLAS:2022ryk} extracts \atau from a shape analysis of the \muonic \pt. 

In summary, an observation of \PGt lepton pair production in ultraperipheral nucleus-nucleus collisions is reported. Events with a final state of one muon and three charged hadrons assumed to be pions are reconstructed from a lead-lead data sample with an integrated luminosity of 404\mubinv collected by the CMS experiment at $\sqrtsNN = 5.02\TeV$ in 2015. The statistical significance of the signal relative to the background-only expectation is far above five standard deviations. The cross section for the \gammagammatautau process, within a fiducial phase space region, is \xsecresult, in agreement with leading-order quantum electrodynamics predictions. Using the measured cross section and its corresponding uncertainties, we estimate a model-dependent value of the anomalous magnetic moment of the \PGt lepton of $\atau=\limit$. This measurement provides a novel experimental probe of the \PGt anomalous magnetic moment using heavy ion collisions at the LHC. 

\begin{acknowledgments}
 We congratulate our colleagues in the CERN accelerator departments for the excellent performance of the LHC and thank the technical and administrative staffs at CERN and at other CMS institutes for their contributions to the success of the CMS effort. In addition, we gratefully acknowledge the computing centers and personnel of the Worldwide LHC Computing Grid and other centers for delivering so effectively the computing infrastructure essential to our analyses. Finally, we acknowledge the enduring support for the construction and operation of the LHC, the CMS detector, and the supporting computing infrastructure provided by the following funding agencies: BMBWF and FWF (Austria); FNRS and FWO (Belgium); CNPq, CAPES, FAPERJ, FAPERGS, and FAPESP (Brazil); MES and BNSF (Bulgaria); CERN; CAS, MoST, and NSFC (China); MINCIENCIAS (Colombia); MSES and CSF (Croatia); RIF (Cyprus); SENESCYT (Ecuador); MoER, ERC PUT and ERDF (Estonia); Academy of Finland, MEC, and HIP (Finland); CEA and CNRS/IN2P3 (France); BMBF, DFG, and HGF (Germany); GSRI (Greece); NKFIH (Hungary); DAE and DST (India); IPM (Iran); SFI (Ireland); INFN (Italy); MSIP and NRF (Republic of Korea); MES (Latvia); LAS (Lithuania); MOE and UM (Malaysia); BUAP, CINVESTAV, CONACYT, LNS, SEP, and UASLP-FAI (Mexico); MOS (Montenegro); MBIE (New Zealand); PAEC (Pakistan); MES and NSC (Poland); FCT (Portugal); MESTD (Serbia); MCIN/AEI and PCTI (Spain); MOSTR (Sri Lanka); Swiss Funding Agencies (Switzerland); MST (Taipei); MHESI and NSTDA (Thailand); TUBITAK and TENMAK (Turkey); NASU (Ukraine); STFC (United Kingdom); DOE and NSF (USA).  
\end{acknowledgments}

\bibliography{auto_generated}

\cleardoublepage \appendix\section{The CMS Collaboration \label{app:collab}}\begin{sloppypar}\hyphenpenalty=5000\widowpenalty=500\clubpenalty=5000\input{HIN-21-009-public-authorlist.tex}\end{sloppypar}
%%% END EDITABLE REGION %%%
% skeleton_end
\end{document}

%% file: HIN-21-009-public-authorlist.tex
\cmsinstitute{Yerevan Physics Institute, Yerevan, Armenia}
{\tolerance=6000
A.~Tumasyan\cmsAuthorMark{1}\cmsorcid{0009-0000-0684-6742}
\par}
\cmsinstitute{Institut f\"{u}r Hochenergiephysik, Vienna, Austria}
{\tolerance=6000
W.~Adam\cmsorcid{0000-0001-9099-4341}, J.W.~Andrejkovic, T.~Bergauer\cmsorcid{0000-0002-5786-0293}, S.~Chatterjee\cmsorcid{0000-0003-2660-0349}, K.~Damanakis\cmsorcid{0000-0001-5389-2872}, M.~Dragicevic\cmsorcid{0000-0003-1967-6783}, A.~Escalante~Del~Valle\cmsorcid{0000-0002-9702-6359}, P.S.~Hussain\cmsorcid{0000-0002-4825-5278}, M.~Jeitler\cmsAuthorMark{2}\cmsorcid{0000-0002-5141-9560}, N.~Krammer\cmsorcid{0000-0002-0548-0985}, L.~Lechner\cmsorcid{0000-0002-3065-1141}, D.~Liko\cmsorcid{0000-0002-3380-473X}, I.~Mikulec\cmsorcid{0000-0003-0385-2746}, P.~Paulitsch, F.M.~Pitters, J.~Schieck\cmsAuthorMark{2}\cmsorcid{0000-0002-1058-8093}, R.~Sch\"{o}fbeck\cmsorcid{0000-0002-2332-8784}, D.~Schwarz\cmsorcid{0000-0002-3821-7331}, S.~Templ\cmsorcid{0000-0003-3137-5692}, W.~Waltenberger\cmsorcid{0000-0002-6215-7228}, C.-E.~Wulz\cmsAuthorMark{2}\cmsorcid{0000-0001-9226-5812}
% ...
\par}
\vskip\cmsinstskip
\dag:~Deceased\\
$^{1}$Also at Yerevan State University, Yerevan, Armenia\\
$^{2}$Also at TU Wien, Vienna, Austria\\
% ...
